\documentclass[10pt,a4paper]{article}
\usepackage[dvips]{color}
\usepackage{epsfig}
\usepackage{amsmath}
\usepackage{graphicx}
\usepackage{authblk}
\usepackage{amssymb,amsmath}
\usepackage{amsmath}

\begin{document}
\title{\bf Higher derivative corrections of $f\left( R \right)$ gravity with varying equation of state in the case of variable $G$ and $\Lambda$}
\author{{M. Khurshudyan$^{a,}$\thanks{Email: martiros.khurshudyan@mpikg.mpg.de}},\hspace{1mm} B. Pourhassan$^{b,}$\thanks{Email: bpourhassan@yahoo.com},\hspace{1mm} A. Pasqua$^{c,}$\thanks{Email: toto.pasqua@gmail.com}\\
$^{a}${\small {\em Max Planck Institute of Colloids and Interfaces, Potsdam-Golm Science Park Am Mühlenberg 1 OT Golm 14476 Potsdam}}\\
$^{b}${\small {\em Department of Physics, Damghan University, Damghan, Iran}}\\
$^{c}${\small {\em Department of Physics, University of Trieste, Via Valerio, 2 34127 Trieste, Italy}}} \maketitle
\begin{abstract}
In this paper, we study three models of $f\left( R \right)$  modified gravity including higher order terms based on different equation of state parameter. We also assume variable $G$ and $\Lambda$. By using numerical analysis, we obtain the behavior of some important cosmological parameters, like the Hubble expansion $H$ parameter and the deceleration parameter $q$. The statefinder diagnostics is also performed for all models.
{\bf Pacs Number(s):} 95.35.+d, 95.85.-e, 98.80.-k
\end{abstract}

\section{Introduction}
The accelerated expansion of universe may be described by dark energy with positive energy and negative pressure [1, 2]. There are several theories to describe the dark energy such as quintessence [3, 4]. Another candidate is Einstein's
cosmological constant which has fine tuning and coincidence problems [5]. There are also
other interesting models to describe the dark energy such as
$k$-essence model [6] and tachyonic model [7]. An interesting
model proposed in order to describe dark energy is the Chaplygin gas (CG) [8, 9] which is not consistent with
observational data [10-13]. Therefore, an extension of
CG model which can unify both dark matter and dark energy [14-16] was proposed and was called generalized Chaplygin gas (GCG). It is also possible to study the possibility of the presence of viscosity in GCG [17-22]. However, observational data ruled out GCG and modified Chaplygin gas (MCG) [23]. Recently, viscous MCG is also suggested and studied [24, 25]. A further extension of CG model is called modified cosmic Chaplygin gas (MCCG) which was proposed recently [26, 27, 28, 29]. Also, various
Chaplygin gas models were studied from the holography point of view [30-32].\\
Research concerning to the
modification matter part of field equations give rise of an
understanding that more complex forms for EoS equation
can be considered. On the other hand, the modification of geometrical part gives rise of different modified
theories like $f(R)$ (with $R$ being the Ricci scalar curvature), $f(T)$ (with $T$ being the torsion scalar), $f(G)$ (with $G$ being the Gauss-Bonnet invariant) etc. Among them, $f(R)$ gravity [33, 34, 35, 36] represents a viable alternative
to dark energy and naturally gives rise to accelerating
singularity-free solutions in early and late cosmic epochs [37, 38, 39]. Indeed, the extended theories of gravity can be considered as a new paradigm to prepare shortcomings of general relativity at IR and UV scales [40]. Moreover, it has been found that the higher-order terms in the gravity Lagrangian are required in order to formulate quantum field theory in curved space-times [41]. These theories agree with the observational data of SNeIa
Hubble diagram [42] and can also give interesting theoretical predictions with respect to the CMBR observational results [43].\\
In this paper we consider a model of $f(R)$ modified gravity with higher derivative corrections based on varying equation of state (EoS) which already was studied, for example, by [44] and with varying $G$ and $\Lambda$. As we know the Einstein equations of general relativity do
not permit any variations in the gravitational constant $G$ and
cosmological constant $\Lambda$ because of the fact that the
Einstein tensor has zero divergence and energy conservation law is
also zero. So, some modifications of Einstein equations are
necessary. This is because, if we simply allow $G$ and $\Lambda$ to
be a variable in Einstein equations, then energy conservation law is
violated. Therefore, the study of the varying $G$ and $\Lambda$ can
be done only through modified field equations and modified
conservation laws [45].\\
In this paper, we suggest three models with different EoS parameters and study cosmological parameters , i.e. Hubble, deceleration and EoS parameters.  In Section 2, we give our motivation to choose special kinds of EoS, and in Section 3 we represent field equations which govern our models. In Section 4, we introduce our models and in Section 5 we summarized our numerical results of each models. In Section 6, statefinder diagnostics performed to obtain more details of our models. Finally in Section 7, we write the conclusions of this paper.

\section{Equation of state}
Let us now consider a different approach to the dark energy EoS. We may think at the EoS as an implicit relation such as \cite{Capozziello}:
\begin{equation}\label{1}
F(P_{x},\rho_{x},H)=0,
\end{equation}
which is not constrained to lead to a linear dependence of $P_{x}$ on $\rho_{x}$. As a particularly example, we will consider the case corresponding to:
\begin{equation}\label{eq:EoS}
(\rho_{x}+P_{x})^{2}-C_{s}\rho_{x}^{2}\left(1-\frac{H_{s}}{H}\right)=0,
\end{equation}
where $C_{s}$ and $H_{s}$ are two positive constants. Eq. (\ref{eq:EoS}) has been
proposed in \cite{Nojiri} where it has been shown that the corresponding
cosmological model presents both Big Bang and Big Rip singularities. It is easy to see that Eq. (\ref{eq:EoS}) has two solutions for $P_{x}$ given by:
\begin{equation}\label{eq:Px}
P_{x}=-\rho_{x}\pm\rho_{x}\sqrt{C_{s} \left (1-\frac{H_{s}}{H} \right )}.
\end{equation}
Motivated by the form of Eq. (3), we would like to investigate a fluid which EoS can we written as:
\begin{equation}\label{eq:P1}
P=\rho+\Omega(t)\rho,
\end{equation}
where $\Omega(t)$ represents  a time-varying parameter. Proposing other modifications from our side, we would like to rewrite Eq. (\ref{eq:P1}) as follow:
\begin{equation}\label{eq:P2}
P=\rho+\Omega(t)\rho^{n}.
\end{equation}
We can consider two possible choices for $\Omega(t)$. In the first case, we consider:
\begin{equation}\label{6}
\Omega(t)=\omega(t)=\omega_{0}+\omega_{1}t \left ( \frac{\dot{H}}{H} + \frac{\dot{G(t)}}{G(t)}\right ),
\end{equation}
where the overdot represents the first time derivative with respect to the time and $G\left(  t \right)$ represents the time dependant gravitational constant.\\
Instead, in the second case, we assume that:
\begin{equation}\label{7}
\Omega(t)=\omega_{1}(t)=\sqrt{A \left (1-\frac{B}{H} \right )}.
\end{equation}
In this paper, we would like to stress that we have three different models for a fluid which governs the background dynamics of the universe in a higher derivative theory of gravity in the presence of time varying gravitational $G(t)$ and $\Lambda$. Within modified theories of gravity we have a hope to solve the problems of dark energy which are originated from general relativity. A gravitational action with higher order term in the scalar curvature $R$ containing a variable $G(t)$ given as:
\begin{equation}\label{eq:Lag}
I = - \int {d^{4}x \sqrt{-g} \left [ \frac{1}{ 16 \pi G(t)} f(R) +L_{m} \right ]},
\end{equation}
where $f(R)$ is a function of $R$ and its higher power including a variable $\Lambda$, $g$ is the determinant of the four dimensional metric and $L_{m}$ represents the matter Lagrangian. Variation of the action given by the Eq. (\ref{eq:Lag}) with respect to $g_{\mu\nu}$ yields to the following equation:
\begin{eqnarray}\label{9}
8\pi G(t) T_{\mu\nu}&=&\frac{1}{2}f(R)g_{\mu\nu}-f_{RR}(R)(\nabla_{\mu}\nabla_{\mu}R-g_{\mu\nu}\nabla^{\mu}\nabla^{\nu}g_{\mu\nu})\nonumber\\
&-&f_{R}(R)R_{\mu\nu}-f_{RRR}(R)(\nabla_{\mu}R\nabla_{\mu}R-\nabla^{\sigma}R \nabla_{\sigma} g_{\mu\nu}),
\end{eqnarray}
where $\nabla_{\mu}$ is the covariant differential operator, $f_{R}(R)$ represents the derivative of $f(R)$ with respect to $R$ and $T_{\mu \nu}$ is the effective energy momentum tensor for matter determined by $L_{m}$.

\section{Field Equations}
Let us consider a higher order gravity as follow \cite{Paul}:
\begin{equation}\label{10}
f(R)=R + \alpha R^{2} -2\Lambda(t).
\end{equation}
It is a well-known fact in the cosmology that the model based on Eq. (10) cannot provide late time acceleration for the present cosmic acceleration but can be used for the inflation in the early universe. The dominance of $\alpha$ term is opposite with the successful expansion history of the universe. Therefore we should discuss about value of $\alpha$ to find appropriate model.\\
By using the Friedmann-Robertson-Walker (FRW) metric for a flat universe, given by,
\begin{equation}\label{eq:s2}
ds^2=-dt^2+a(t)^2\left(dr^{2}+r^{2}d\Omega^{2}\right),
\end{equation}
we can obtain the following expression \cite{Paul}:
\begin{equation}\label{eq:Fridmmanvlambda}
H^{2}-6\alpha(2H\ddot{H}-\dot{H}^{2}+6\dot{H}H^{2})=\frac{8\pi G(t)\rho}{3}+\frac{\Lambda(t)}{3},
\end{equation}
where $d\Omega^{2}=d\theta^{2}+\sin^{2}\theta d\phi^{2}$ is the angular part of the metric and $a(t)$
represents the scale factor of the universe, which gives information about the expansion of the universe. $\theta$ and $\phi$ parameters are
the usual azimuthal and polar angles of spherical coordinates, with $0\leq\theta\leq\pi$ and $0\leq\phi<2\pi$. The coordinates ($t, r,
\theta, \phi$) are called co-moving coordinates.\\
The energy conservation is given by:
\begin{equation}\label{eq:conservation}
\dot{\rho}+3H(\rho+P)=- \left( \frac{\dot{G}}{G}\rho +\frac{\dot{\Lambda}}{8\pi G}\right ),
\end{equation}
where $\rho$ and $P$ are the energy density and pressure of the perfect fluid, respectively.
Modified theories of gravity like $f(R)$ theories give us possibility to find a natural representation and introduction of the dark energy into theory. Comparing Eq. (\ref{eq:Fridmmanvlambda}) with the field equations in the general relativity, we can associate the term $6\alpha(2H\ddot{H}-\dot{H}^{2}+6\dot{H}H^{2})$ to the energy density of dark energy. Therefore, the type of dark energy and dynamics of the universe depends on the form of $f(R)$ which will be considered. The type of the work which we would like to consider in this paper assumed an existence of an effective fluid controlling the dynamics of the universe composed non interacting dark energy (from $f(R)$) and a fluid from our assumptions. In order to
obtain viable cosmological model, one can consider perfect fluid as a source of matter in the framework of higher derivative gravity. It is known that higher order gravity with suitable counter terms and $\Lambda$ added to the Einstein-Hilbert action,
one gets a perturbation theory which is well behaved, formally renormalizable and
asymptotically free \cite{Paul}.\\
In forthcoming articles we will consider different interactions considered in literature. As there is not interaction for mater we have the following expression for the conservation equation:
\begin{equation}\label{eq:drho}
\dot{\rho}+3H(\rho+P)=0.
\end{equation}
Therefore, for the dynamics of $G(t)$ we will have the following expression:
\begin{equation}\label{dotg}
\dot{G}+\frac{\dot{\Lambda}(t)}{8\pi \rho}=0.
\end{equation}
In the next Section we introduce our model and specific form of $\Lambda$ which assumed in this paper.
\section{The models}
In order to perform dynamical analysis of the universe we also assume that the form of $\Lambda$ is given. In order to avoid future confusions and to do not complicate the writing and the analysis of the work, we assumed that the form of an effective $\Lambda$
is given and it is a summation of the well known and well studied forms of $\Lambda$, namely we take:
\begin{equation}\label{eq:lambda}
\Lambda(t)=t^{-2}+\gamma \rho +\dot{H},
\end{equation}
where $\gamma$ is a positive constants which controls the contribution of energy density of the matter in $\Lambda$. Therefore, for the dynamics of $G$ we will obtain, from Eq. (\ref{dotg})
\begin{equation}\label{17}
\dot{G} +\frac{ \gamma \dot{\rho} + \ddot{H} -2t^{-3} }{8\pi \rho}=0.
\end{equation}
We can write Eq. (\ref{eq:drho}) in an other form, i.e.:
\begin{equation}\label{18}
\dot{\rho}_{i}+3H\rho_{i}(1+\omega_{i})=0,
\end{equation}
where the terms $\omega_{i}$ indicates the EoS parameters of the fluids and defined as $P_{i}/\rho_{i}$.\\
In this paper, we consider three different models as follow.
\subsection{Model 1}
In the first model, motivated by equation (5), we choose the following expression for the pressure [46],
\begin{equation}\label{19}
P=\rho+\omega(t)\rho^{n},
\end{equation}
where $\omega(t)$ is given by Eq. (6). Therefore, the equation of state is obtained as follow,
\begin{equation}\label{20}
\omega=1+\omega(t)\rho^{n-1}.
\end{equation}
In Ref. [49], barotropic fluid with quadratic EoS introduced. Now, the relation (19) may be extended version of EoS to higher order. In the case of $n=2$ and $\omega(t)=Const$, the quadratic EoS parameter of the Ref. [49] is recovered.
\subsection{Model 2}
Another choice, which is our second model, described by the following pressure [46, 48],
\begin{equation}\label{21}
P=\rho-\omega_{1}(t)\rho^{n/2},
\end{equation}
where $\omega_{1}(t)$ is given by Eq. (7).
Therefore, the equation of state obtained as follow:
\begin{equation}\label{22}
\omega=1-\omega_{1}(t)\rho^{n/2-1}.
\end{equation}
This model is also similar to the previous model with some modifications which affect our final numerical results.
\subsection{Model 3}
Our third model is based on modified Chaplygin gas EoS and it is given by:
\begin{equation}\label{23}
P=A\rho-\frac{B}{\rho^{n}},
\end{equation}
where $A$, $B$ and $n$ are positive constants.
Therefore, the equation of state obtained as the follow,
\begin{equation}\label{24}
\omega=A-\frac{B}{\rho^{n+1}}.
\end{equation}
This case is also a special form of general equation (5) and we expected that it yields to more appropriate results than the previous models.
\section{Numerical results}
In this Section, we present results of our numerical analysis of the models we are dealing with. Comparing our results with observational data helps to choose one of them as the best model.
\subsection{Model 1}
The Hubble expansion parameter $H$ for this model is illustrated in Fig. 1. The left plot shows that increasing $n$ decreases value of the Hubble parameter, but right plot shows that increasing $\omega_{0}$ increases value of the Hubble parameter. We can see that Hubble parameter is decreasing function of time at the initial stage but is increasing function at the late time. This is unusual behavior, because we expected that the Hubble parameter should yields to a constant ($H\approx70$ at present stage).\\
The deceleration parameter $q$ for this model illustrated in  Fig. 2. The left plot shows that increasing $n$ increases net value of the deceleration parameter, but right plot shows that increasing $\omega_{0}$ decreases net value of the deceleration parameter. This is completely opposite of the Hubble parameter. Therefore, we can see that deceleration parameter is increasing function of time initially and is decreasing function at the late time. As we can see the value of the deceleration parameter is negative in all times, so there is no acceleration to deceleration phase transition. Observational data give us $q>-1$ and $\Lambda$CDM model predict $q\rightarrow-1$. Therefore this model ruled out by observational data. However still we can study other parameter which may be interesting from theoretical point of view.\\
The EoS parameter of the model 1 is illustrated in Fig. 3 with variation of $\gamma$. We can see $P/\rho>0$ which reduces to +1 at the late time. This also disagrees with observations which suggest $\omega\rightarrow-1$.\\
Finally, the evolution of $\dot{G}/G$ presented in Fig. 4 which is an increasing function of time and has not upper bound [50], therefore we can not accept the model 1 as a good model.\\\\
\begin{figure}[h!]
 \begin{center}$
 \begin{array}{cccc}
\includegraphics[width=50 mm]{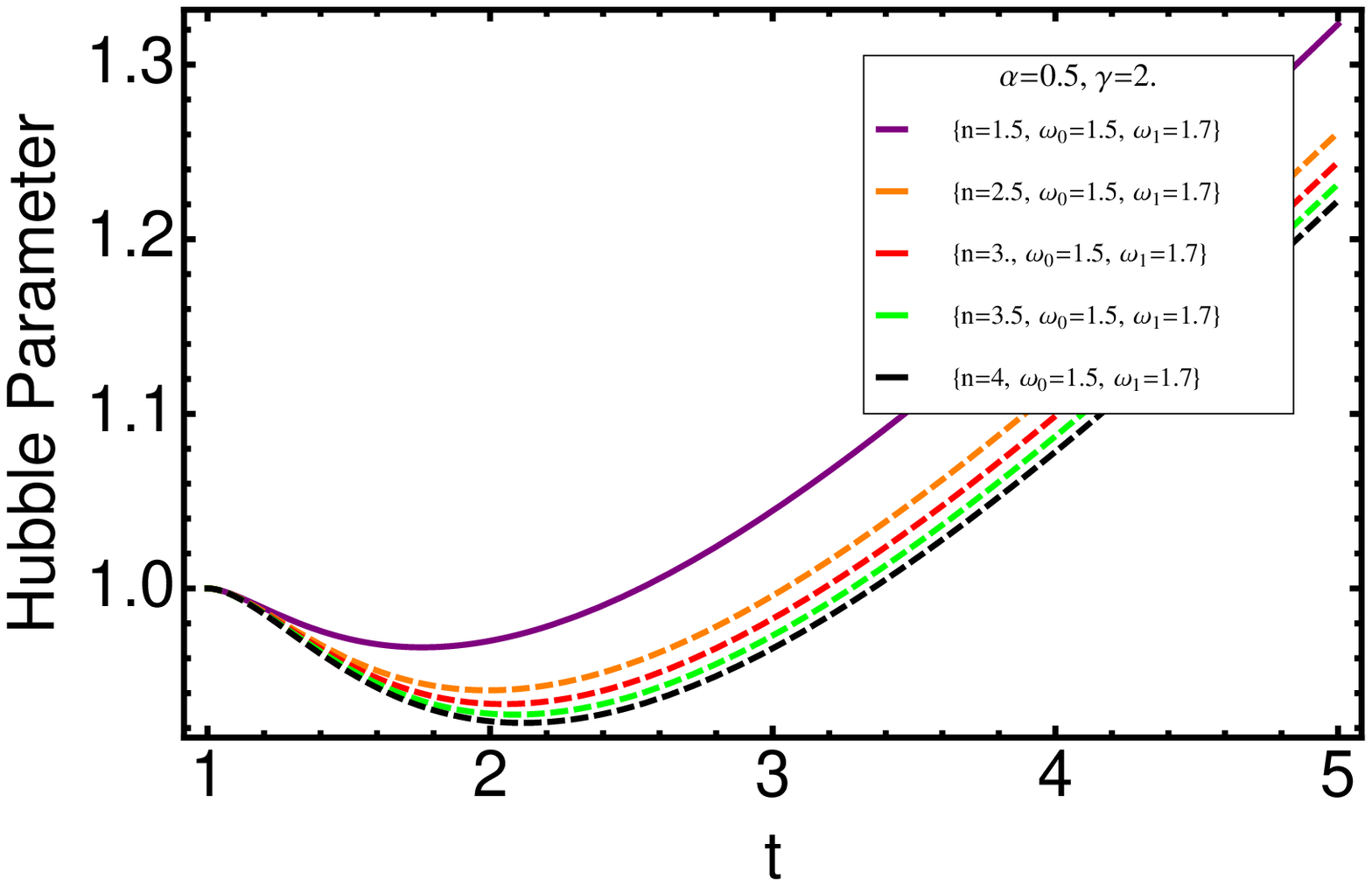}&
\includegraphics[width=50 mm]{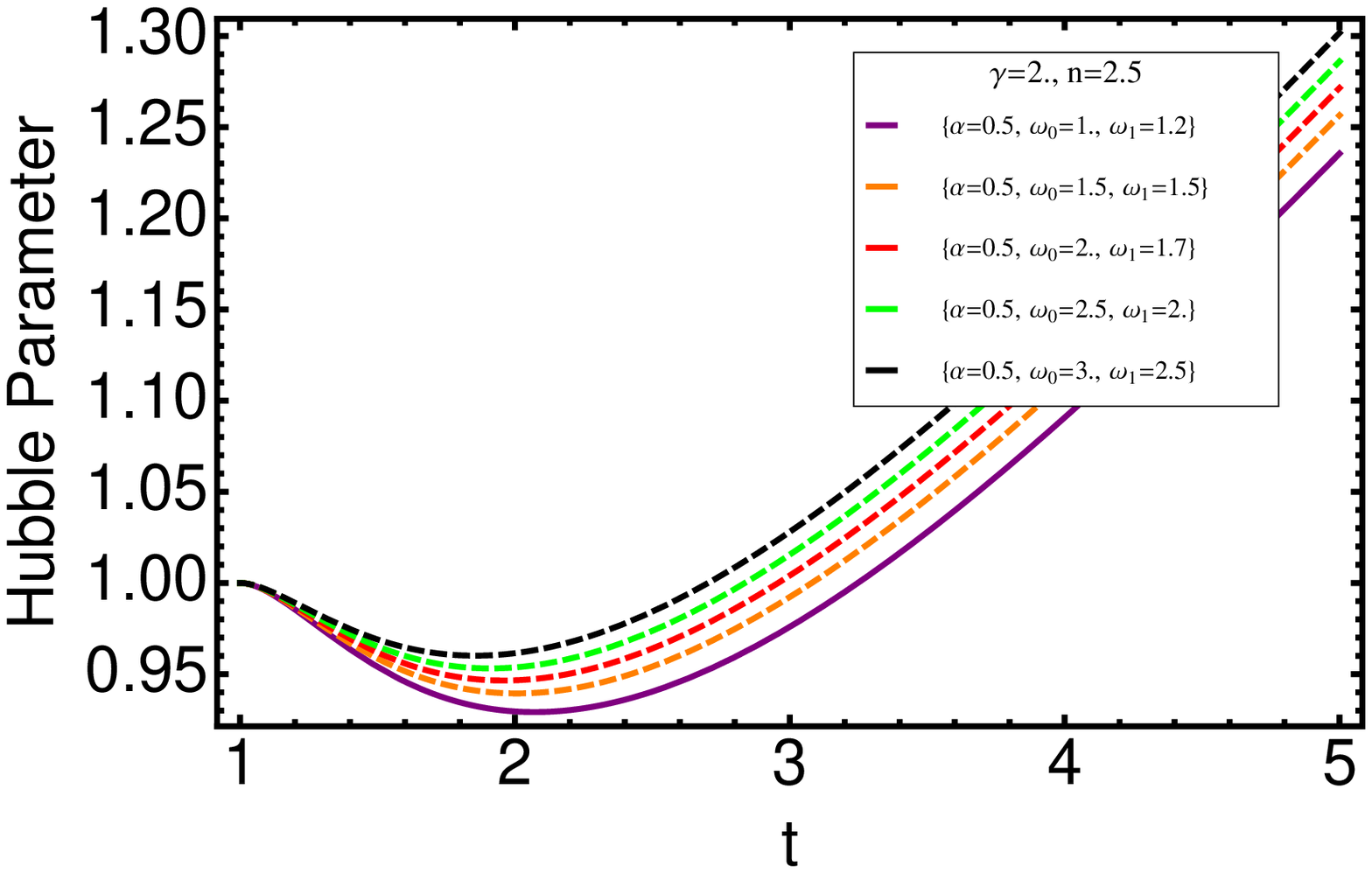}
 \end{array}$
 \end{center}
\caption{Behavior of $H$ against $t$. Model 1}
 \label{fig:1}
\end{figure}

\begin{figure}[h!]
 \begin{center}$
 \begin{array}{cccc}
\includegraphics[width=50 mm]{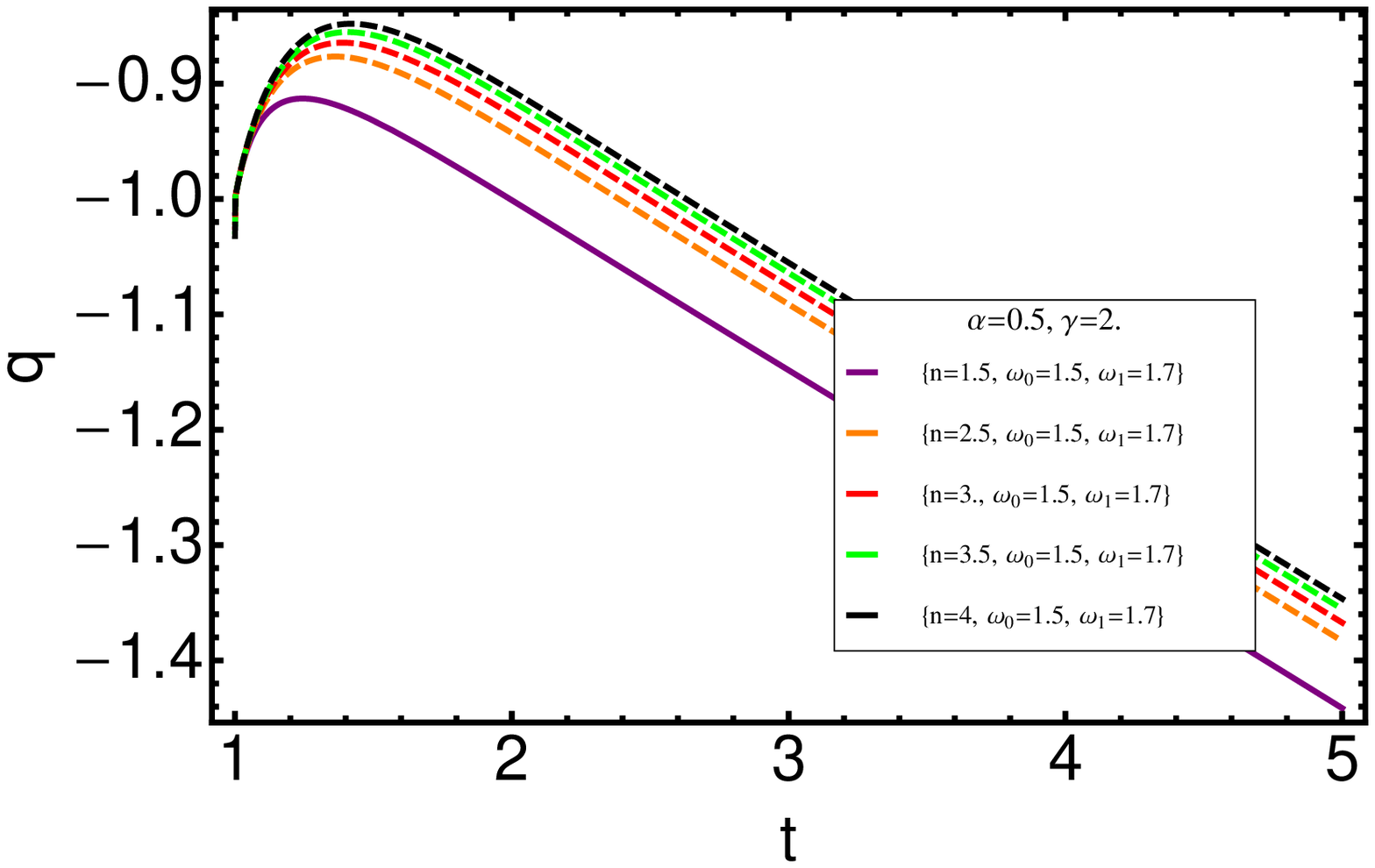}&
\includegraphics[width=50 mm]{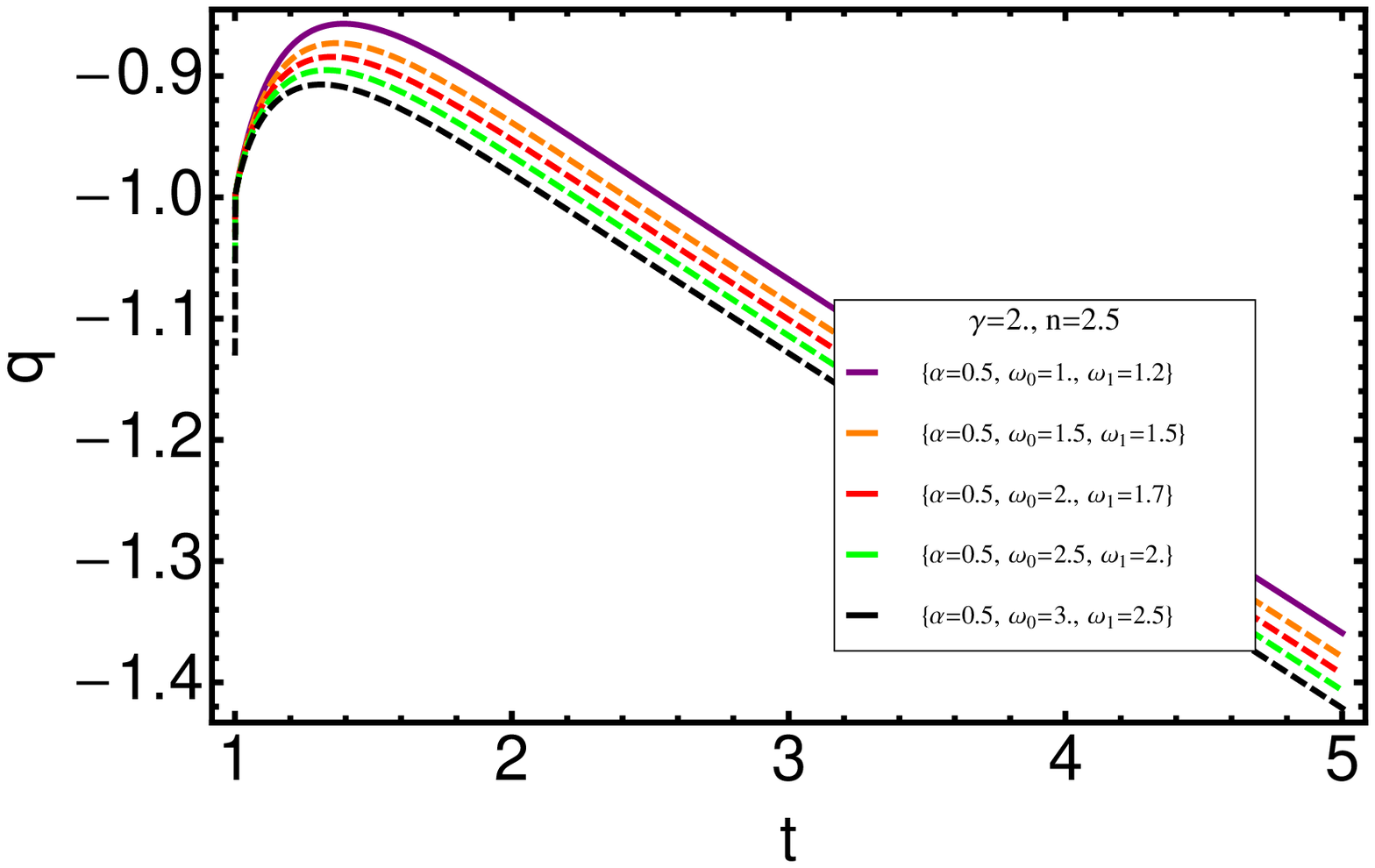}
 \end{array}$
 \end{center}
\caption{Behavior of $q$ against $t$. Model 1}
 \label{fig:2}
\end{figure}

\begin{figure}[h!]
 \begin{center}$
 \begin{array}{cccc}
\includegraphics[width=50 mm]{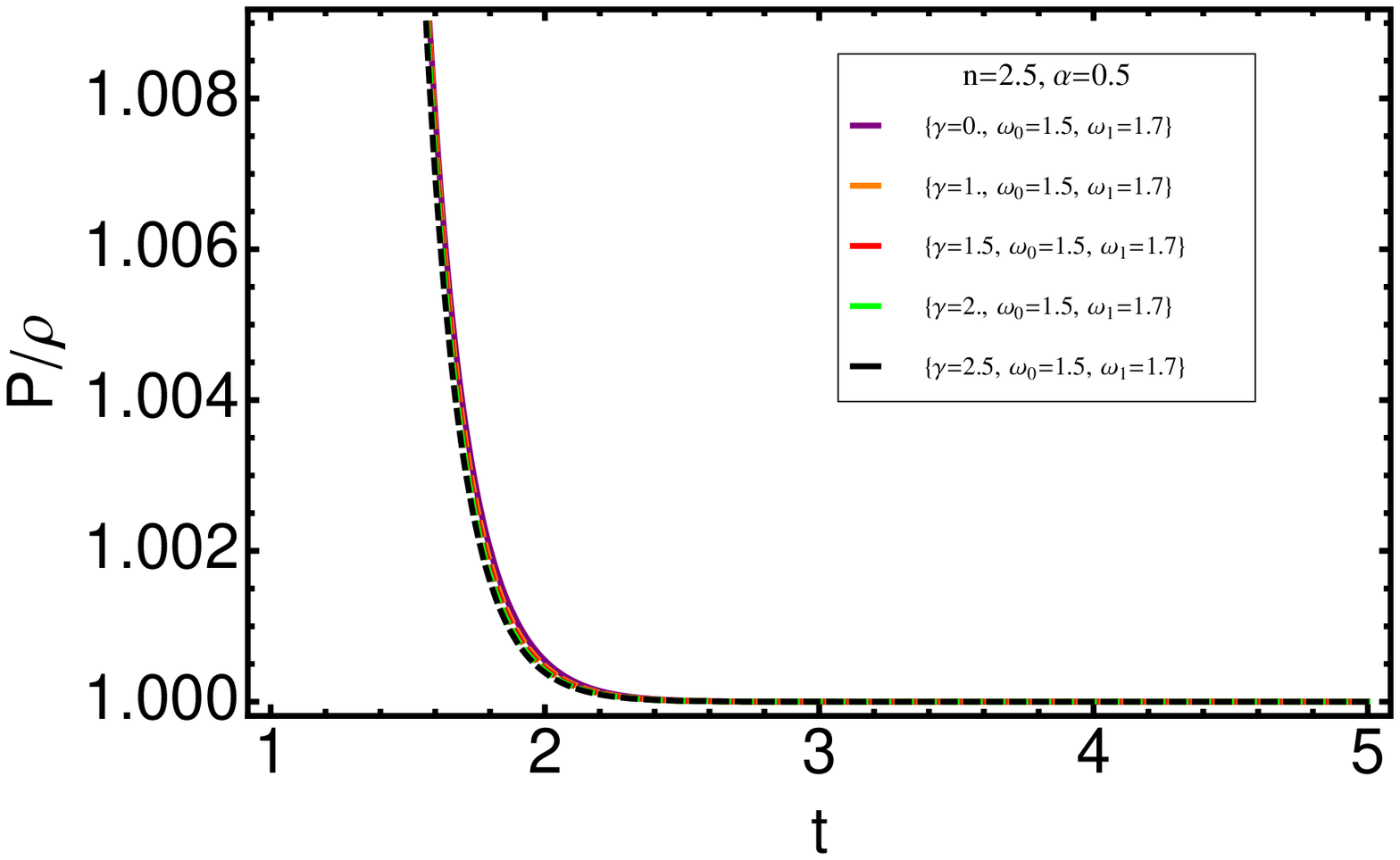}\\
 \end{array}$
 \end{center}
\caption{Behavior of $\omega$ against $t$. Model 1}
 \label{fig:3}
\end{figure}

\begin{figure}[h!]
 \begin{center}$
 \begin{array}{cccc}
\includegraphics[width=50 mm]{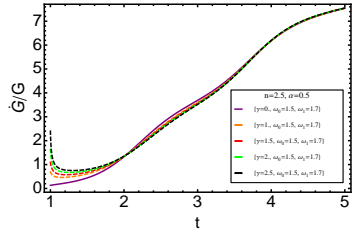}\\
 \end{array}$
 \end{center}
\caption{Behavior of $\dot{G}/G$ against $t$. Model 1}
 \label{fig:4}
\end{figure}

\subsection{Model 2}
As we have shown, Model 1 disagrees with observational data, therefore we used the EoS given in Eq.  (22) instead of that given in Eq.  (20) and constructed the model 2. The Hubble parameter of the model 2 is presented by plots of Fig. 5. The left plot shows that increasing $\alpha$ increases value of the Hubble parameter, but right plot shows that increasing $\gamma$ decreases value of the Hubble parameter. We can see that Hubble parameter is decreasing function of time as expected. Therefore we can see that lower values of $\alpha$ and higher values of $\gamma$ (approximately $\alpha=1$ and $\gamma=3$) yields to $H\approx70$ (0.7 in our scale) which is approximately the current value of the Hubble expansion parameter.\\
The deceleration parameter $q$ of the model 2 is illustrated in Fig. 6. The left plot shows that increasing the value of $\alpha$ decreases the net value of the deceleration parameter $q$, but right plot shows that increasing $\gamma$ increases net value of the deceleration parameter. We can see that deceleration parameter is increasing function of time initially and is decreasing function at the late time to yields a constant value. As we can see, the value of the deceleration parameter is negative at all times, so as the previous model, there is no acceleration to deceleration phase transition. However, we obtained $q>-1$ which agrees with some current observational data.\\
The EoS parameter of the model 2 is illustrated in Fig. 7 for different values of $\alpha$. It is shown that increasing the value of $\alpha$ decreases the net value of $\omega$. As we expected for a dark energy model, the EoS parameter is obtained as a negative parameter.\\
Finally, the evolution of $\dot{G}/G$ is presented in Fig. 8, which has a singular point and finally yields to a constant satisfied constraint of from Viking landers on Mars data [51] ($\dot{G}/G\leq6$ in our scale) and also constraints from the pulsar systems  PSR B1913+16 and PSR B1855+09 [52] ($\dot{G}/G\leq9$ in our scale).\\\\

\begin{figure}[h!]
 \begin{center}$
 \begin{array}{cccc}
\includegraphics[width=50 mm]{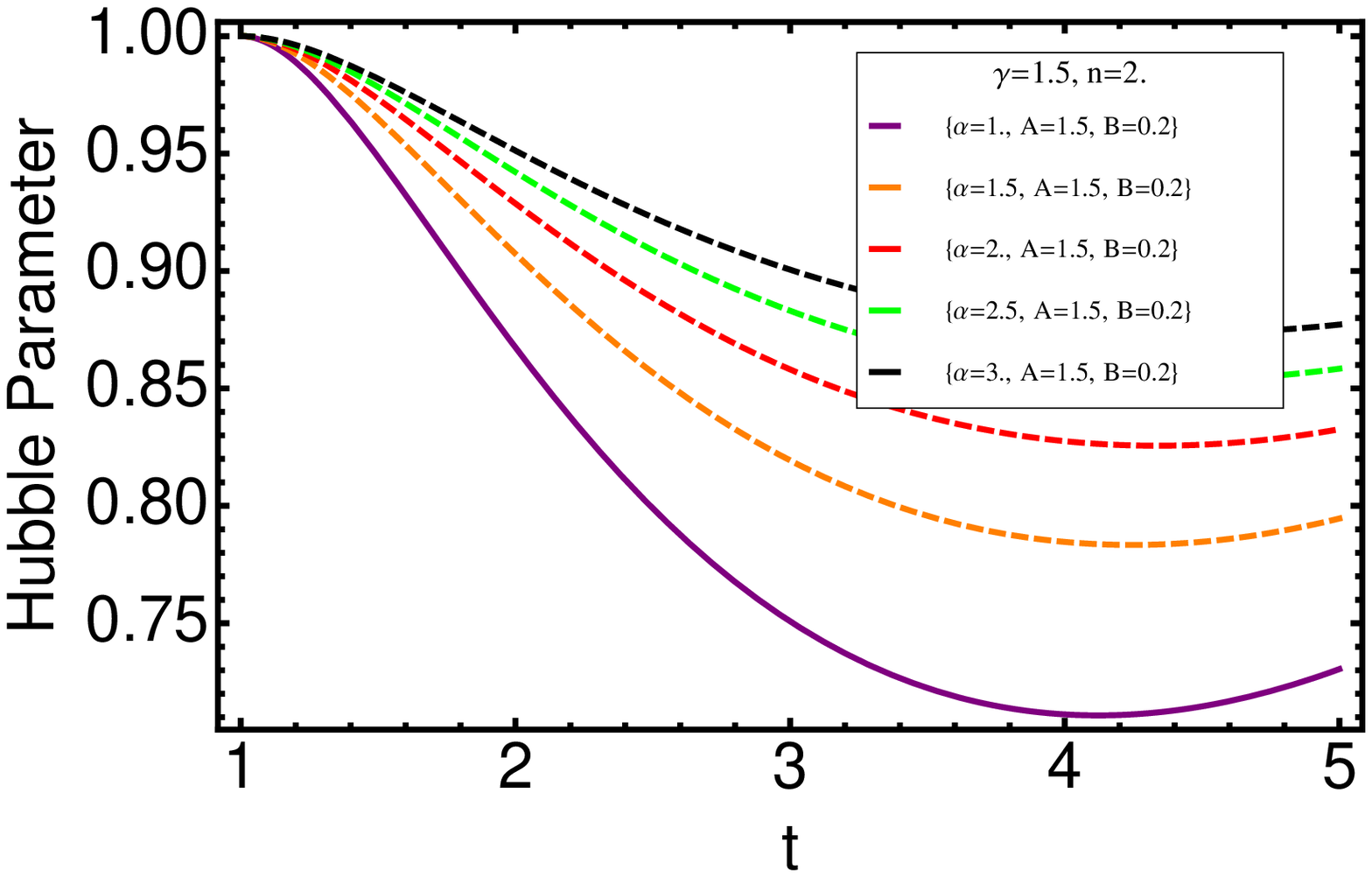} &
\includegraphics[width=50 mm]{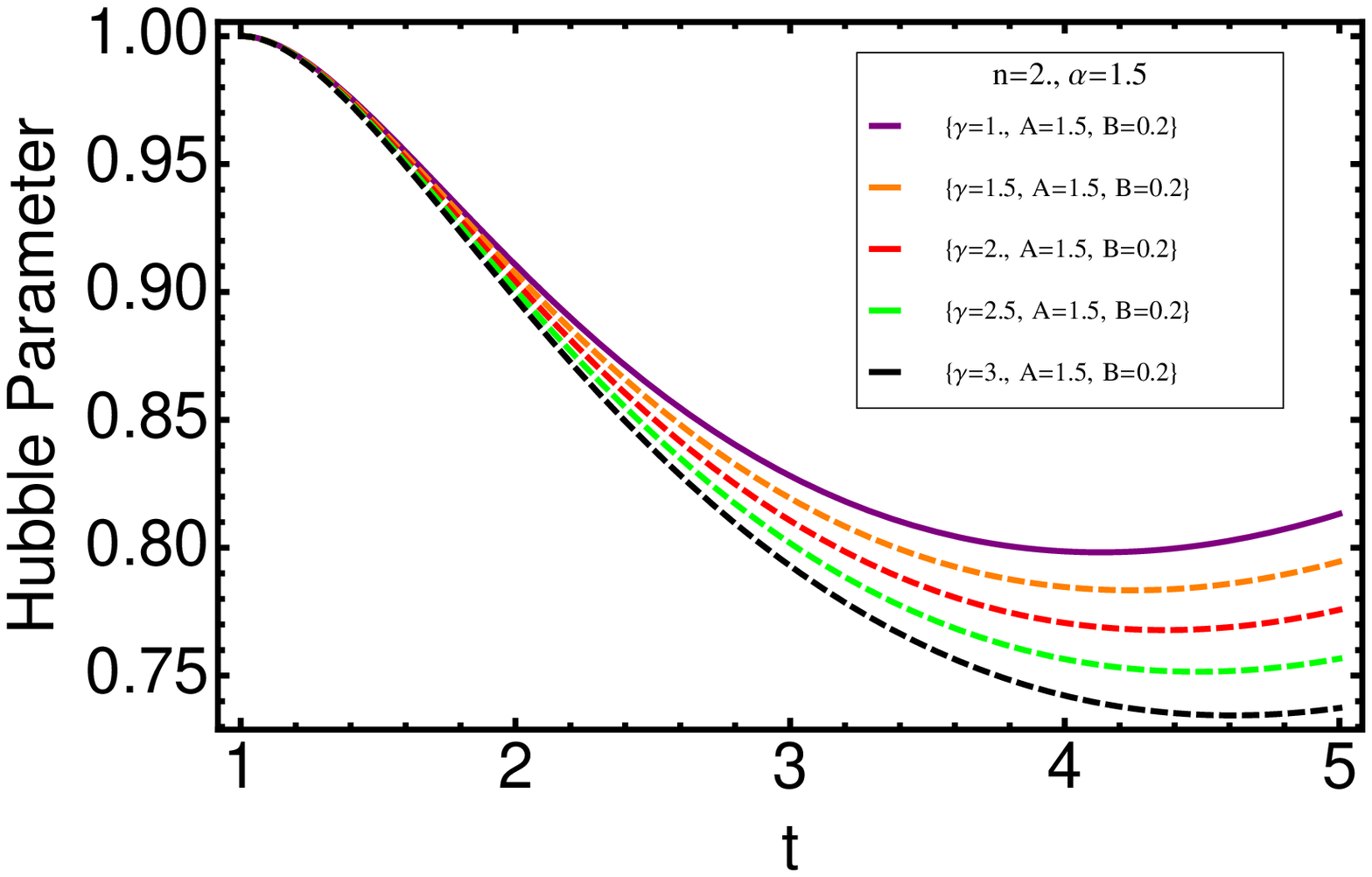}
 \end{array}$
 \end{center}
\caption{Behavior of $H$ against $t$. Model 2}
 \label{fig:5}
\end{figure}

\begin{figure}[h!]
 \begin{center}$
 \begin{array}{cccc}
\includegraphics[width=50 mm]{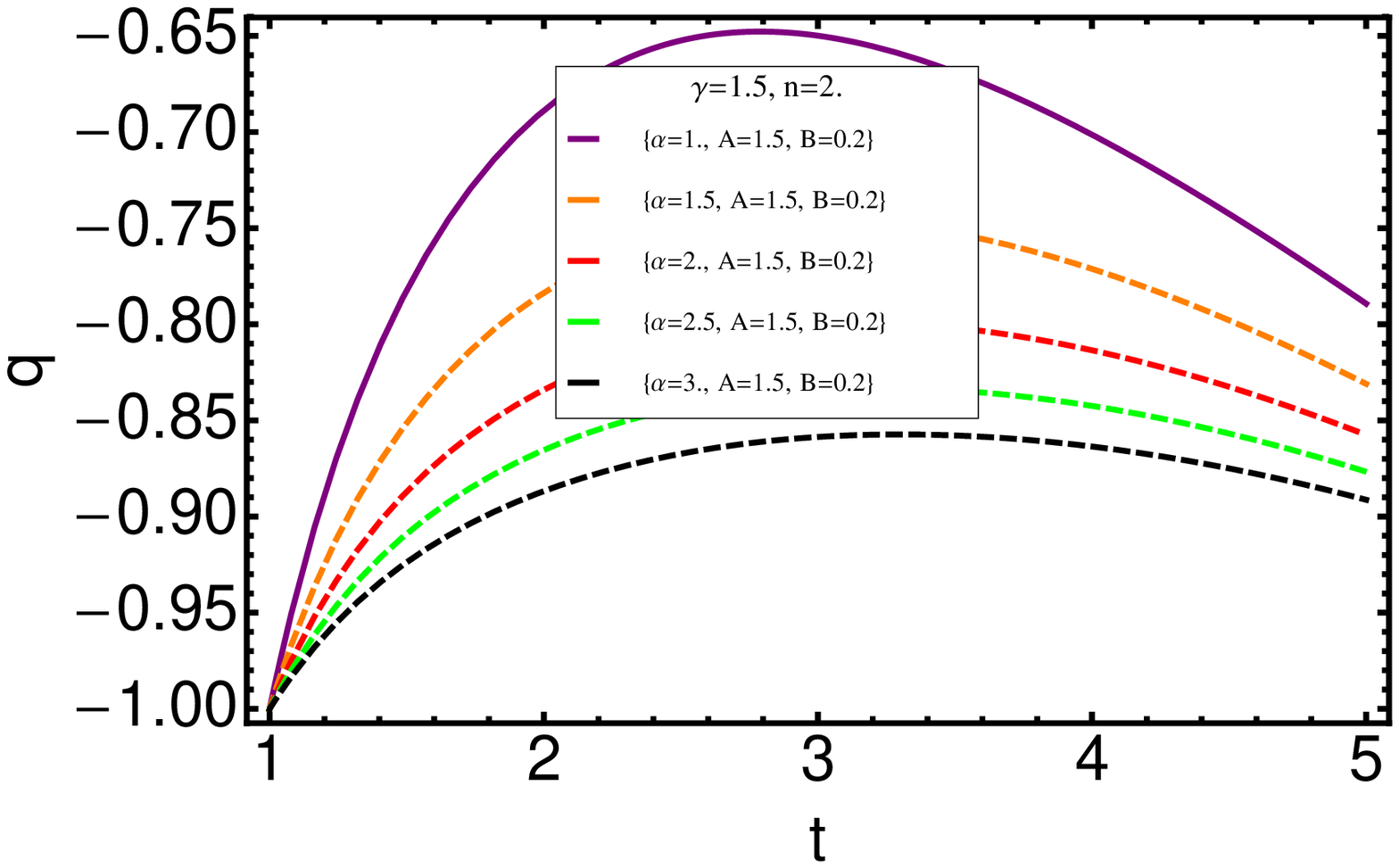} &
\includegraphics[width=50 mm]{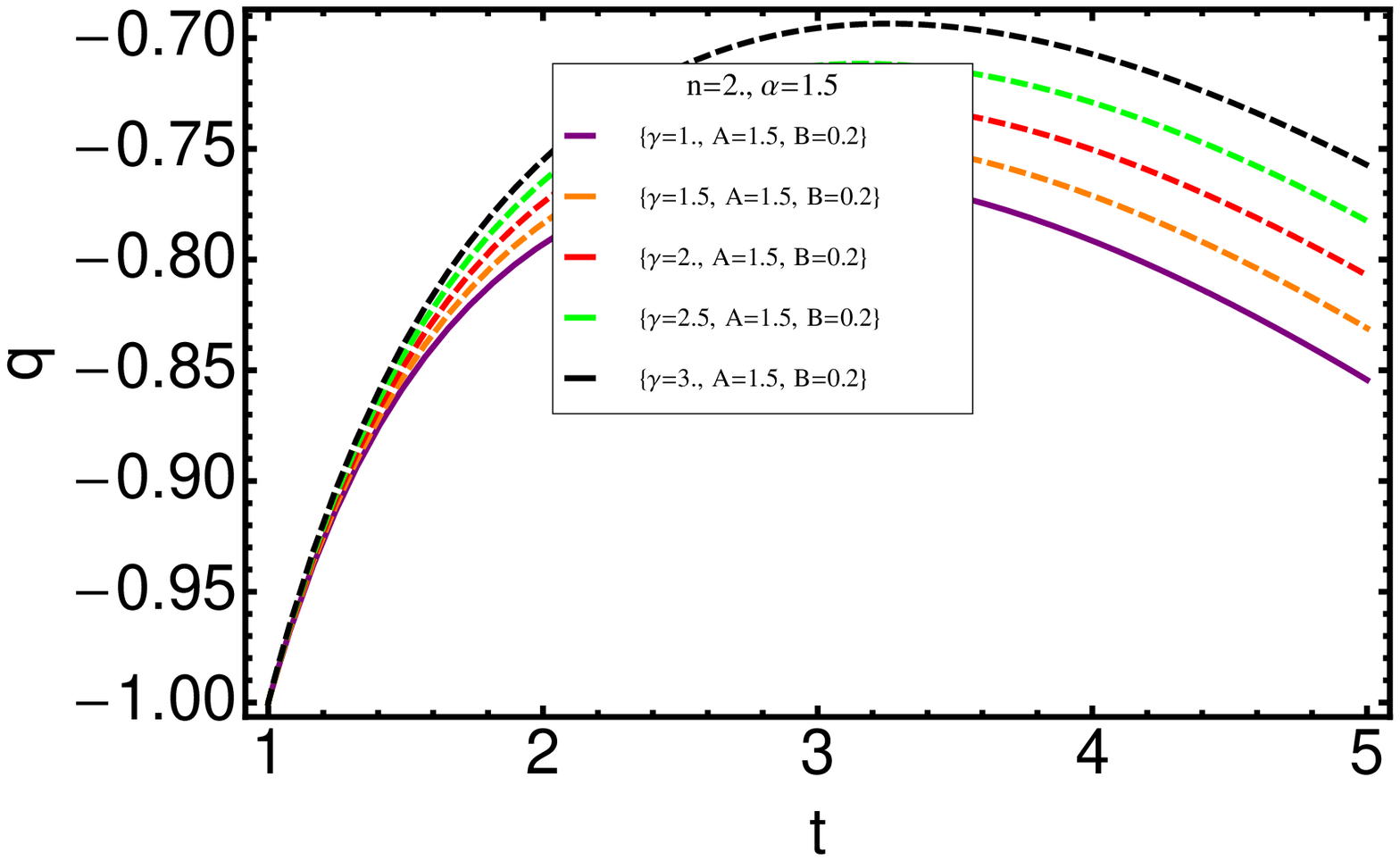}
 \end{array}$
 \end{center}
\caption{Behavior of $q$ against $t$. Model 2}
 \label{fig:6}
\end{figure}

\begin{figure}[h!]
 \begin{center}$
 \begin{array}{cccc}
\includegraphics[width=50 mm]{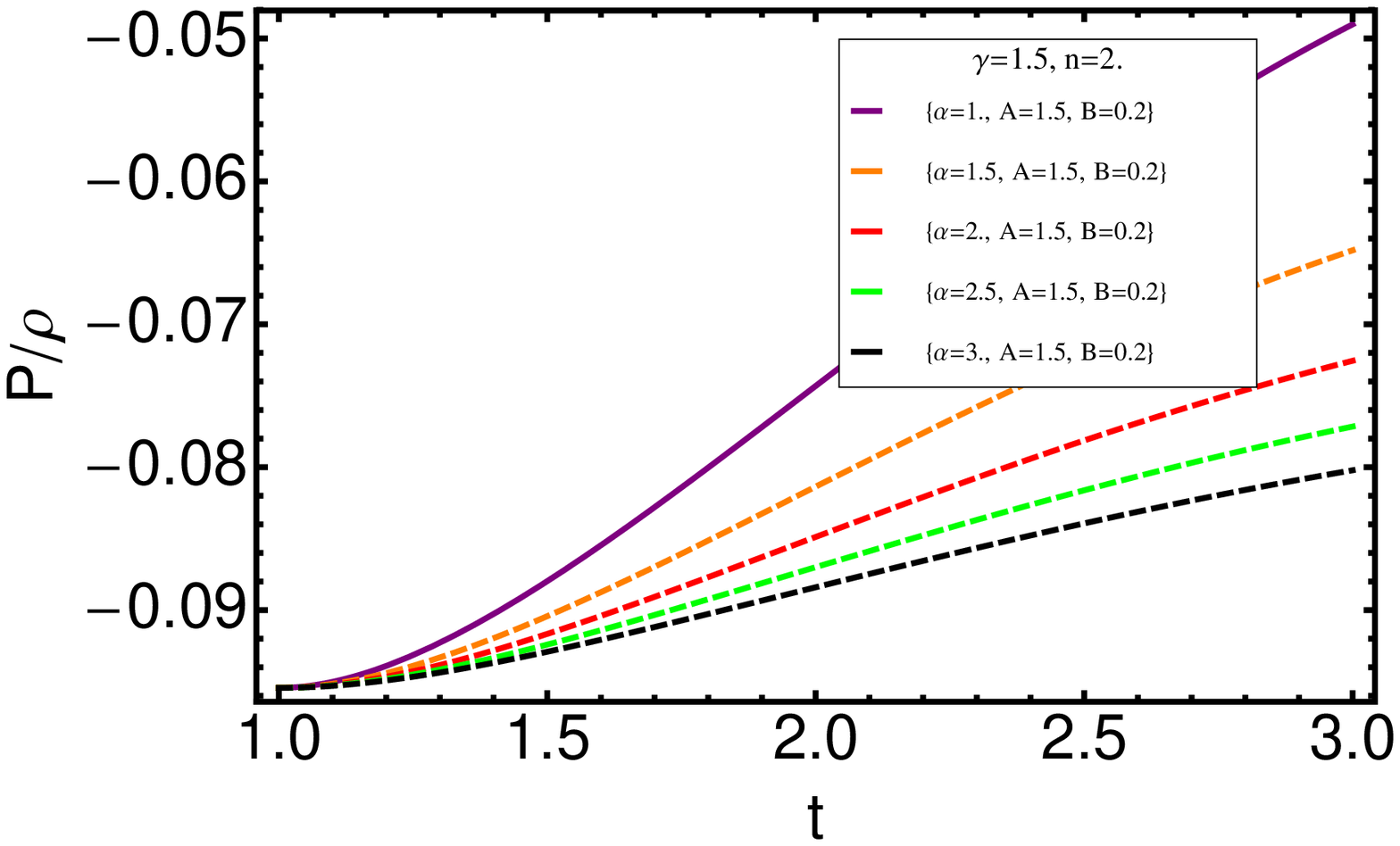}
 \end{array}$
 \end{center}
\caption{Behavior of $\omega$ against $t$. Model 2}
 \label{fig:7}
\end{figure}

\begin{figure}[h!]
 \begin{center}$
 \begin{array}{cccc}
\includegraphics[width=50 mm]{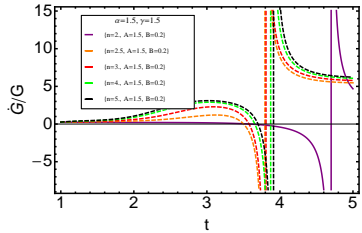}
 \end{array}$
 \end{center}
\caption{Behavior of $\dot{G}/G$ against $t$. Model 2}
 \label{fig:8}
\end{figure}

\subsection{Model 3}
The Hubble expansion parameter $H$ of the model 3 is presented in the plots of Fig. 9. It is  obtained a decreasing function of the time and yields to a constant value lower than current observational data. We also find that, similar to the previous models, increasing $\alpha$ increases value of the Hubble parameter.\\
The deceleration parameter $q$ of the model 3 is illustrated in Fig. 10. We find that increasing $\gamma$ increases the net value of the deceleration parameter $q$. We can see that deceleration parameter is increasing function of time, also deceleration to acceleration phase transition may be found in this model while we expect acceleration to deceleration phase transition. This opposite behavior may be cause of ruling out the modified Chaplygin gas cosmology [53]. However, we expected $q>-1$ which agrees with some current observational data.\\
The EoS parameter  of the model 3 is illustrated in Fig. 11 for varying $n$. It is shown that increasing $n$ decreases net value of $\omega$. We can see that value of $\omega$ is positive initially and then take negative value to yields $\omega\rightarrow-1$ which agrees with current observational data.\\
Finally, the evolution of $\dot{G}/G$ is presented in Fig. 12: we can see that it has a critical point denoted with $t_{c}$. For the $\dot{G}/G\leq t_{c}$ it is decreasing function of time. Then, for $\dot{G}/G> t_{c}$ any dependence on parameters are inverse and its value yields to a constant in agreement with observational data [50].\\\\

\begin{figure}[h!]
 \begin{center}$
 \begin{array}{cccc}
\includegraphics[width=50 mm]{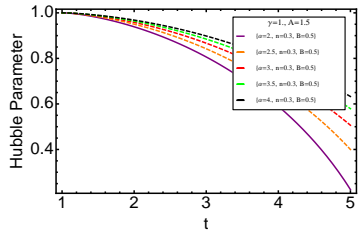}
 \end{array}$
 \end{center}
\caption{Behavior of $H$ against $t$. Model 3}
 \label{fig:9}
\end{figure}

\begin{figure}[h!]
 \begin{center}$
 \begin{array}{cccc}
\includegraphics[width=50 mm]{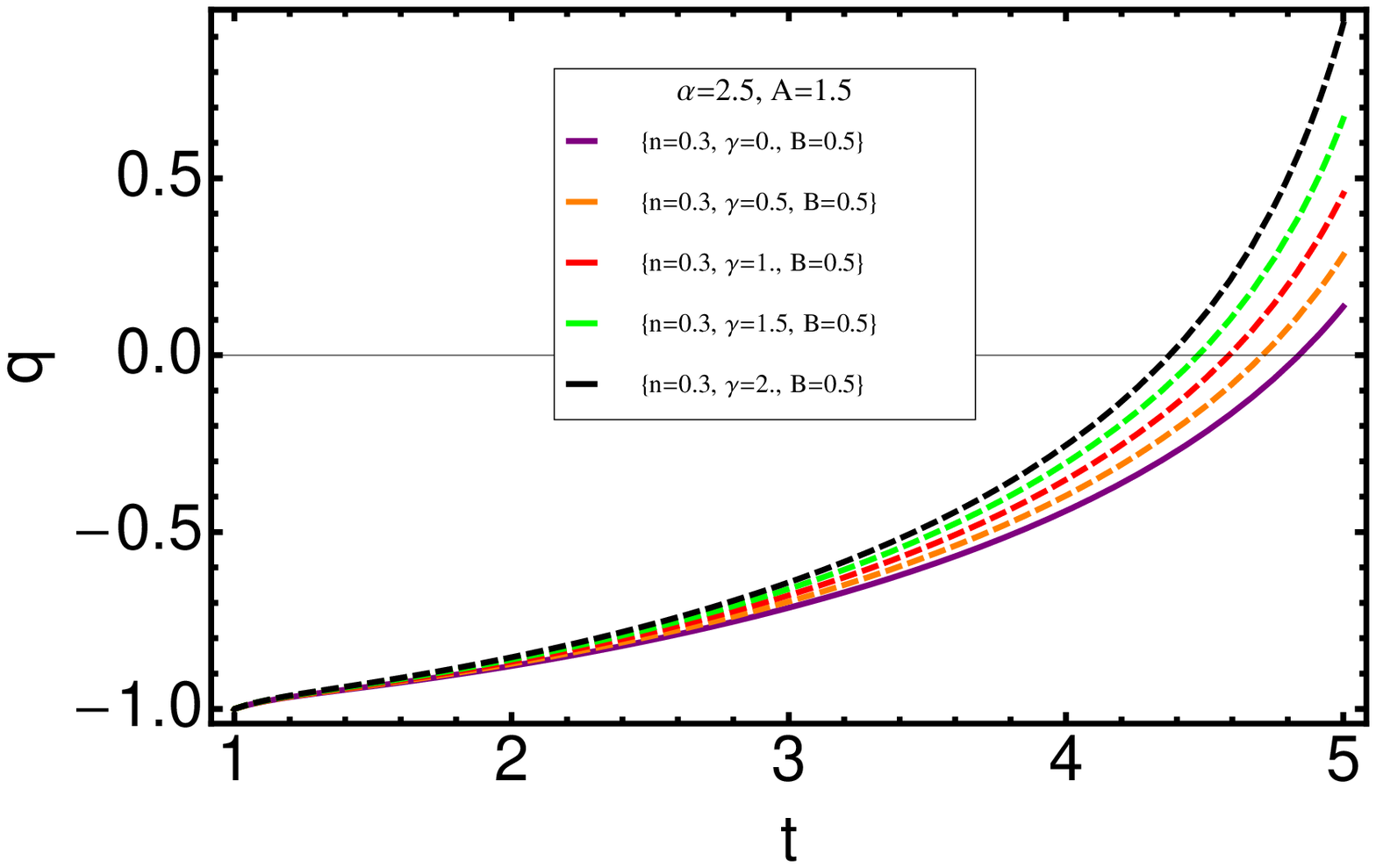}
 \end{array}$
 \end{center}
\caption{Behavior of $q$ against $t$. Model 3}
 \label{fig:10}
\end{figure}

\begin{figure}[h!]
 \begin{center}$
 \begin{array}{cccc}
\includegraphics[width=50 mm]{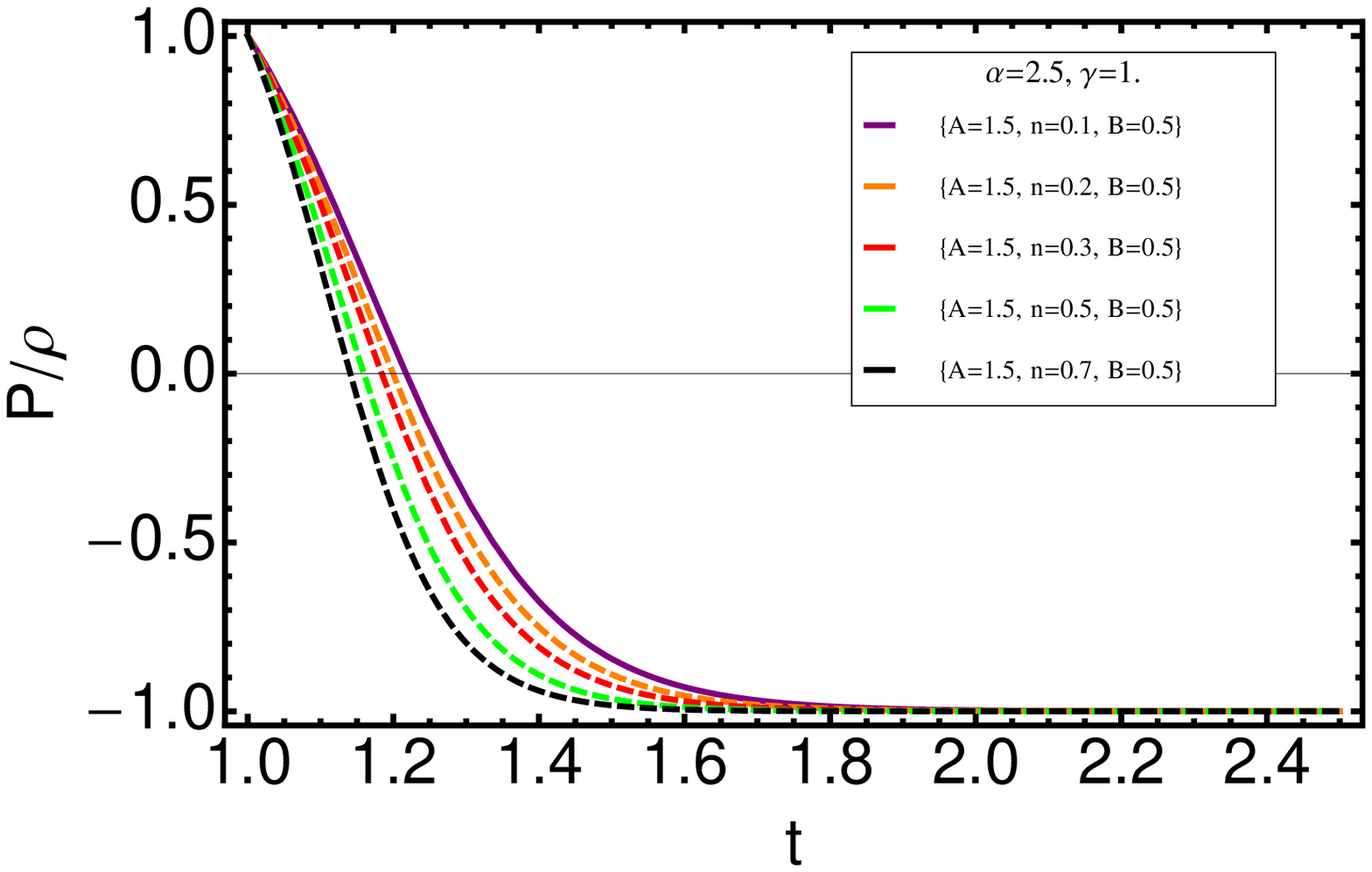}
 \end{array}$
 \end{center}
\caption{Behavior of $\omega$ against $t$. Model 3}
 \label{fig:11}
\end{figure}

\begin{figure}[h!]
 \begin{center}$
 \begin{array}{cccc}
\includegraphics[width=50 mm]{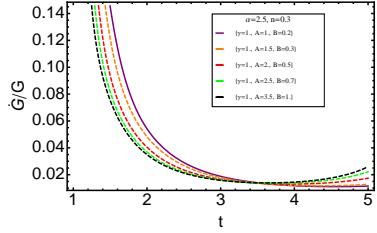}
 \end{array}$
 \end{center}
\caption{Behavior of $\dot{G}/G$ against $t$. Model 3}
 \label{fig:12}
\end{figure}

\section{Statefinder diagnostics}
In the framework of general relativity, dark energy can explain the present cosmic acceleration. Except cosmological constant, there are many others candidates of dark energy (quintom, quintessence, brane, modified gravity etc.) proposed in scientific literature. One of the properties of dark energy is that it is model dependent and, in order to differentiate different models of dark energy, a sensitive diagnostic tool is needed. The Hubble parameter $H$ and the deceleration parameter $q$ are very important quantities which can describe the geometric properties of the universe. Since $\dot{a}>0$, hence $H>0$ implies an expansion of the universe. Moreover, $\ddot{a}>0$, which means $q<0$, indicates an accelerated expansion of the universe. Therefore, various dark energy models give $H>0$ and $q<0$, then  they can not provide enough evidence to differentiate the more accurate cosmological observational data and the more general models of dark energy. For this aim, we need  higher order of time derivative of scale factor and geometrical tool. Sahni \emph{et.al} \cite{Sahni} proposed geometrical statefinder diagnostic tool, based on dimensionless parameters $\left\{r, s\right\}$ which are functions of the scale factor $a$ and its higher order time derivatives. These parameters are defined as follow:
\begin{equation}\label{eq:statefinder}
r=\frac{1}{H^{3}}\frac{\dddot{a}}{a} ~~~~~~~~~~~~
s=\frac{r-1}{3(q-\frac{1}{2})}.
\end{equation}
Results of our numerical analysis are presented in the Fig. 13. We can see that our first and third models completely agree with the results of the Refs. [54, 55, 56]. However model 2 has also acceptable behavior.
\begin{figure}[h!]
 \begin{center}$
 \begin{array}{cccc}
\includegraphics[width=55 mm]{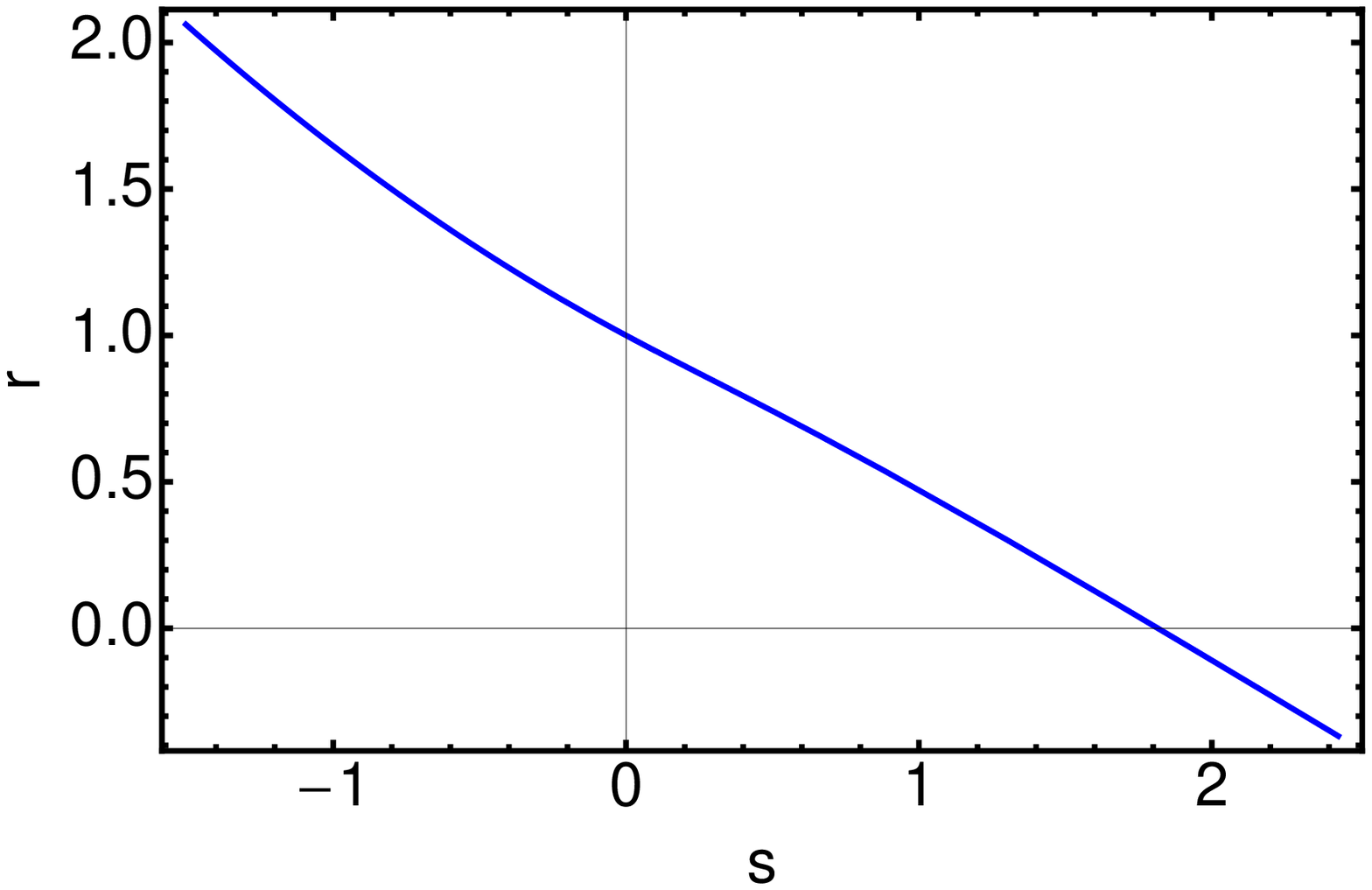} &
\includegraphics[width=45 mm]{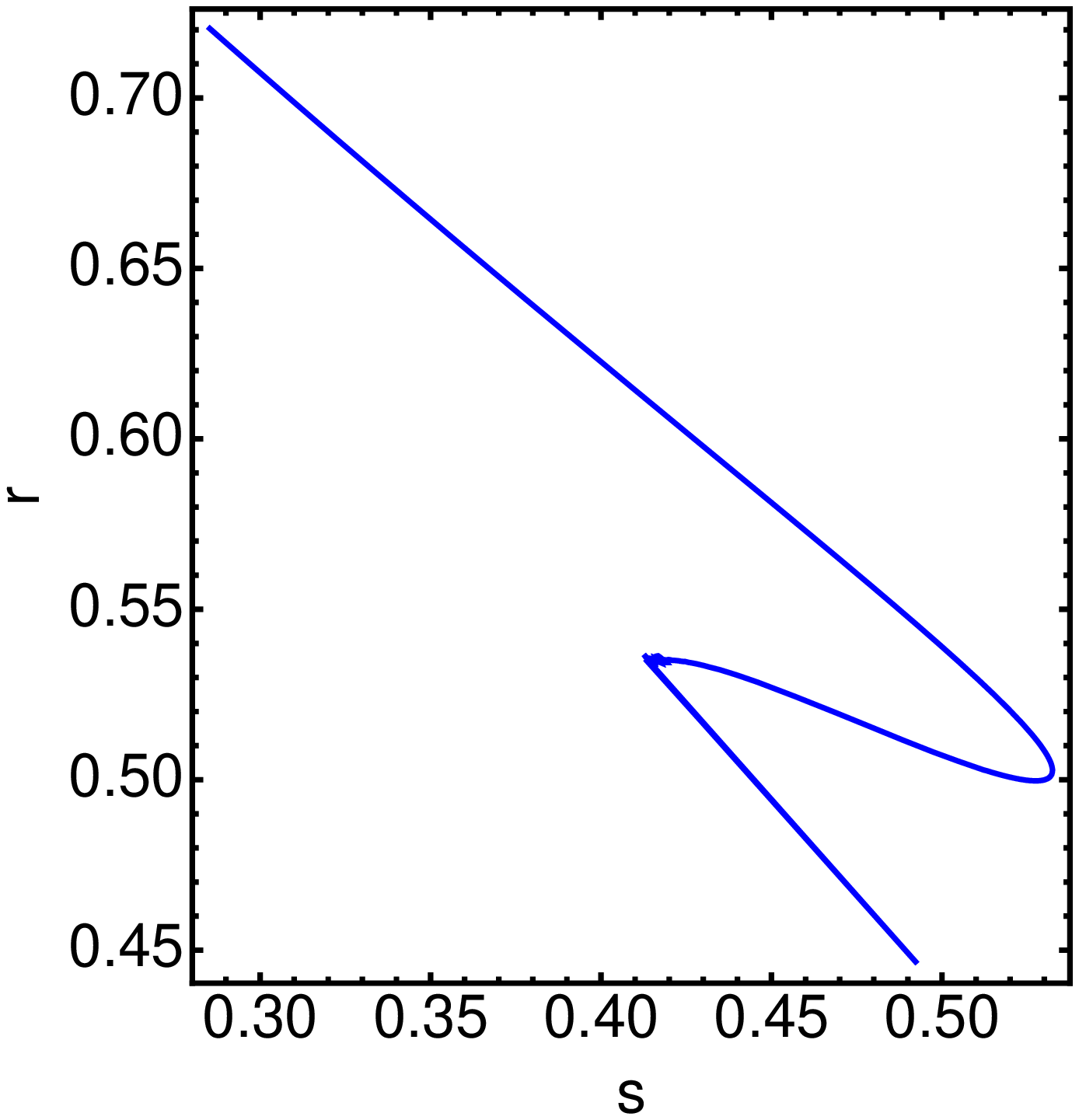}&
\includegraphics[width=55 mm]{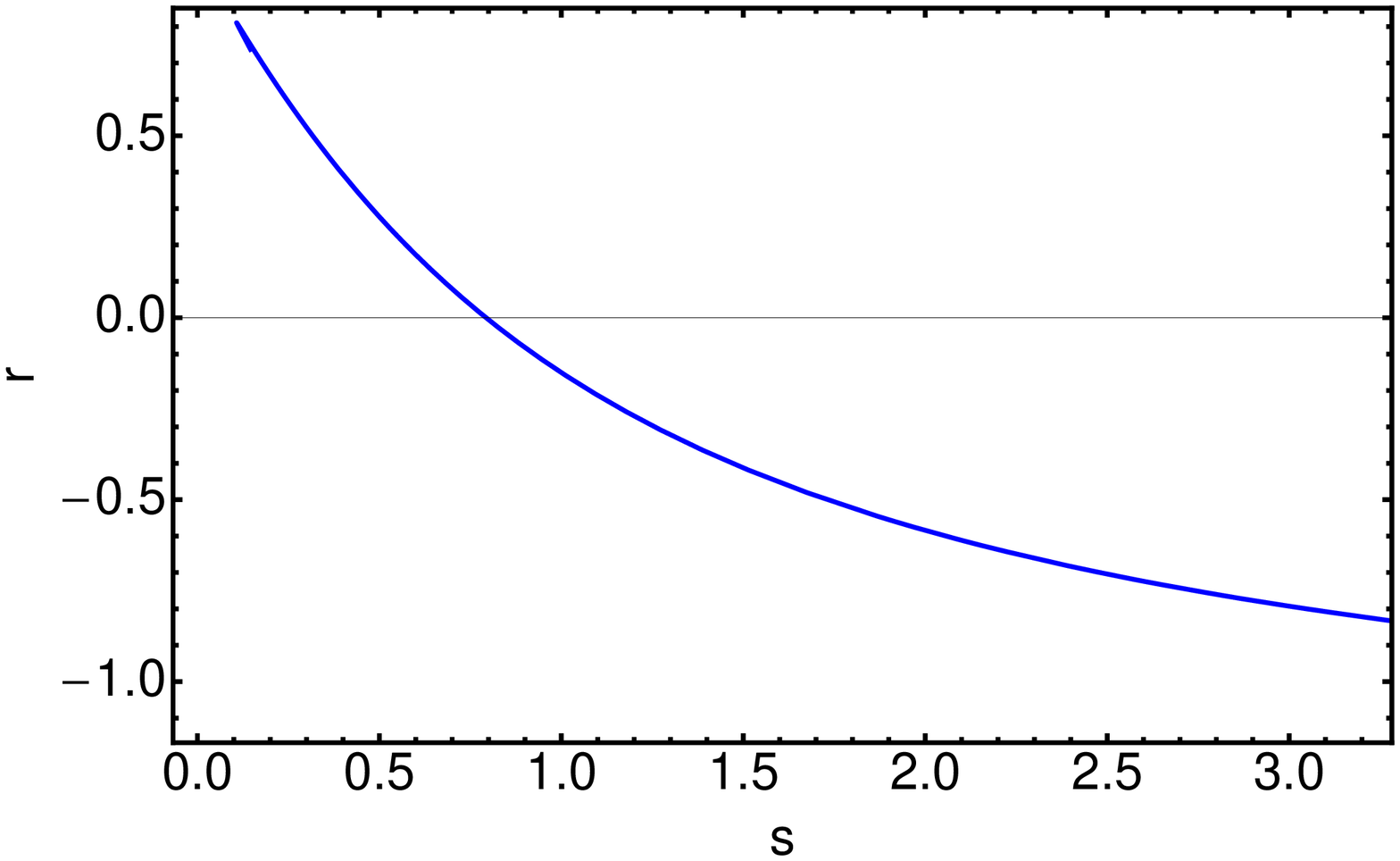}
 \end{array}$
 \end{center}
\caption{$r-s$. Model 1: $\alpha=0.5$, $n=2.5$, $\gamma=1.5$, $\omega_{0}=1.5$ and $\omega_{1}=1.7$. Model 2:  $\alpha=1.5$, $n=2.0$, $\gamma=2.0$, $A=1.5$ and $B=0.2$. Model 3: $\alpha=2.5$, $n=0.3$, $\gamma=0.5$, $A=1.5$ and $B=0.5$}
 \label{fig:13}
\end{figure}

\section{Conclusion}
In this paper, we suggested three different models of dark energy with variable gravitational constant $G$ and $\Lambda$ in the framework of higher order $f\left(R\right)$ gravity. We considered $\Lambda$ as a combination of $t$, $\rho$ and $\dot{H}$ functions, and investigated the effect of higher order terms on the cosmological parameters. Difference of our models is based on EoS parameters. Our numerical analysis shown that effect of higher order terms is increasing Hubble expansion parameter while decreasing deceleration and EoS parameters. Our first and second models motivated by higher order barotropic fluid EoS, but we concluded that the first model is not coincide with some observational data while the second model may agree with current observations. Our third model is based on modified Chaplygin gas EoS. Although some works suggest ruling out of this model but we found some agreements with observational data. However we suggest our second model as the best model of this paper. An important point is value of $\alpha$ parameter which is obtained $\alpha=0.5$ for the first model, $\alpha=1.5$ for the second model and $\alpha=2.5$ for the third model. These may agree with the fact that value of $\alpha$ should be small \cite{55,56}.

\end{document}